\documentclass[twocolumn,showpacs,preprintnumbers,amsmath,amssymb]{revtex4}

\usepackage{graphicx}
\usepackage{dcolumn}
\usepackage{bm}

\begin{document}


\title{Impact of CLAS and COMPASS data on Polarized
Parton Densities and Higher Twist }

\author{Elliot Leader}
\email{e.leader@imperial.ac.uk} \affiliation{Imperial College,
Prince Consort Road, London SW7 2BW, England.
}%

\author{Aleksander V. Sidorov}
\email{sidorov@theor.jinr.ru} \affiliation{ Bogoliubov Theoretical
Laboratory Joint Institute for Nuclear Research 141980 Dubna,
Russia.
}%

\author{Dimiter B. Stamenov}
\email{stamenov@inrne.bas.bg}
\affiliation{ Institute for Nuclear Research and Nuclear Energy \\
Bulgarian Academy of Sciences
Blvd. Tsarigradsko Chaussee 72, Sofia 1784, Bulgaria}%
\date{March 28, 2007}

\begin{abstract}

We have re-analyzed the world data on inclusive polarized DIS
including the very precise CLAS proton and deuteron data, as well
as the latest COMPASS data on the asymmetry $A_1^d$, and have
studied the impact of these data on polarized parton densities and
higher twist effects. We demonstrate that the low $Q^2$ CLAS data
improve essentially our knowledge of higher twist corrections to
the spin structure function $g_1$, while the large $Q^2$ COMPASS
data influence mainly the strange quark density. In our new
analysis we find that a negative polarized gluon density, or one
that changes sign as a function of $x$, cannot be ruled out on the
basis of the present DIS data.
\end{abstract}

\maketitle

\section{Introduction}

The European Muon Collaboration (EMC) experiment \cite{EMC} at
CERN found that a surprisingly small fraction of the proton spin
is carried by the spin of the quarks. This observation was a big
challenge to our understanding of the partonic spin structure of
the nucleon, {\it i.e.}, how the nucleon spin is built up out from
the intrinsic spin and orbital angular momentum of its
constituents, quarks and gluons. Since that time substantial
efforts, both experimental and theoretical, have been made to
answer this question. Our present knowledge about the spin
structure of the nucleon comes mainly from polarized inclusive and
semi-inclusive DIS experiments at SLAC, CERN, DESY and JLab,
polarized proton-proton collisions at RHIC and polarized
photoproduction experiments. One of the important and best studied
aspects of this knowledge is the determination of the longitudinal
polarized parton densities in QCD and their first moments
\cite{groups,LSS05}, which correspond to the spins carried by the
quarks and gluons in the nucleon.

One of the features of polarized DIS is that a lot of the present
data are in the preasymptotic region ($Q^2 \sim 1-5~\rm
GeV^2,~4~\rm GeV^2 < W^2 < 10~\rm GeV^2$). This is especially the
case for the experiments performed at the Jefferson Laboratory. As
was shown in \cite{LSS_HT}, to confront correctly the QCD
predictions to the experimental data including the preasymptotic
region, the {\it non-perturbative} higher twist (powers in
$1/Q^2$) corrections to the nucleon spin structure functions have
to be taken into account too.

In this paper we study the impact of the recent very precise CLAS
\cite{CLAS06} and COMPASS \cite{COMPASS06} inclusive polarized DIS
data on the determination of both the longitudinal polarized
parton densities (PPD) in the nucleon and the higher twist (HT)
effects. These experiments give important information about the
nucleon structure in quite different kinematic regions. While the
CLAS data entirely belong to the preasymptotic region and as one
can expect they should mainly influence the higher twist effects,
the COMPASS data on the spin asymmetry $A_1^d$ are large $Q^2$
data and they should affect mainly the polarized parton densities.
In addition, due to COMPASS measurements we have for the first
time accurate data at small $x~ (0.004 < x < 0.015)$, where the
behaviour of the spin structure function $g_1^d$ should be more
sensitive to the sign of the gluon polarization.

\section{Next to leading QCD analysis of the data}

In QCD the spin structure function $g_1$ has the following form
($Q^2 >> \Lambda^2$):
\begin{equation}
g_1(x, Q^2) = g_1(x, Q^2)_{\rm LT} + g_1(x, Q^2)_{\rm HT}~,
\label{g1QCD}
\end{equation}
where "LT" denotes the leading twist ($\tau=2$) contribution to
$g_1$, while "HT" denotes the contribution to $g_1$ arising from
QCD operators of higher twist, namely $\tau \geq 3$. In Eq.
(\ref{g1QCD}) (the nucleon target label N is dropped)
\begin{eqnarray}
g_1(x, Q^2)_{\rm LT}&=& g_1(x, Q^2)_{\rm pQCD}  + h^{\rm TMC}(x,
Q^2)/Q^2\nonumber\\
&&+ {\cal O}(M^4/Q^4)~, \label{g1LT}
\end{eqnarray}
where $g_1(x, Q^2)_{\rm pQCD}$ is the well known (logarithmic in
$Q^2$) NLO pQCD contribution
\begin{eqnarray}
g_1(x,Q^2)_{\rm pQCD}&=&{1\over 2}\sum _{q} ^{N_f}e_{q}^2 [(\Delta
q +\Delta\bar{q})\otimes (1 + {\alpha_s(Q^2)\over 2\pi}\delta C_q)
\nonumber\\
&&+{\alpha_s(Q^2)\over 2\pi}\Delta G\otimes {\delta C_G\over
N_f}], \label{g1partons}
\end{eqnarray}
and $h^{\rm TMC}(x, Q^2)$ are the calculable kinematic target mass
corrections \cite{TMC}, which effectively belong to the LT term.
In Eq. (\ref{g1partons}), $\Delta q(x,Q^2), \Delta\bar{q}(x,Q^2)$
and $\Delta G(x,Q^2)$ are quark, anti-quark and gluon polarized
densities in the proton, which evolve in $Q^2$ according to the
spin-dependent NLO DGLAP equations. $\delta C(x)_{q,G}$ are the
NLO spin-dependent Wilson coefficient functions and the symbol
$\otimes$ denotes the usual convolution in Bjorken $x$ space. $\rm
N_f$ is the number of active flavors ($\rm N_f=3$ in our
analysis). In addition to the LT contribution, the dynamical
higher twist effects
\begin{equation}
g_1(x, Q^2)_{\rm HT}= h(x, Q^2)/Q^2 + {\cal O}(\Lambda^4/Q^4)~,
\label{HTQCD}
\end{equation}
must be taken into account at low $Q^2$. The latter are
non-perturbative effects and cannot be calculated in a model
independent way. That is why we prefer to extract them directly
from the experimental data. The method used to extract
simultaneously the polarized parton densities and higher twist
corrections to $g_1$ is described in \cite{LSS_HT}. According to
this method, the $g_1/F_1$ and $A_1(\approx g_1/F_1)$ data have
been fitted using the experimental data for the unpolarized
structure function $F_1(x,Q^2)$
\begin{equation}
\left[{g_1(x,Q^2)\over F_1(x, Q^2)}\right]_{exp}~\Leftrightarrow~
{{g_1(x,Q^2)_{\rm LT}+h(x)/Q^2}\over F_1(x,Q^2)_{exp}}~.
\label{g1F2Rht}
\end{equation}
\vskip 0.4cm As usual, $F_1$ is replaced by its expression in
terms of the usually extracted from unpolarized DIS experiments
$F_2$ and $R$ and phenomenological parametrizations of the
experimental data for $F_2(x,Q^2)$ \cite{NMC} and the ratio
$R(x,Q^2)$ of the longitudinal to transverse $\gamma N$
cross-sections \cite{R1998} are used. Note that such a procedure
is equivalent to a fit to $(g_1)_{exp}$, but it is more precise
than the fit to the $g_1$ data themselves actually presented by
the experimental groups because here the $g_1$ data are extracted
in the same way for all of the data sets. Note also, that in our
analysis the logarithmic $Q^2$ dependence of $h(x, Q^2)$ in Eq.
(\ref{g1F2Rht}), which is not known in QCD, is neglected. Compared
to the principal $1/Q^2$ dependence it is expected to be small and
the accuracy of the present data does not allow its determination.
Therefore, the extracted from the data values of $h(x)$ correspond
to the mean $Q^2$ for each $x$-bean (see Table II and the
discussion below).

As in our previous analyses, for the input NLO polarized parton
densities at $Q^2_0=1~GeV^2$ we have adopted a simple
parametrization
\begin{eqnarray}
\nonumber
x\Delta u_v(x,Q^2_0)&=&\eta_u A_ux^{a_u}xu_v(x,Q^2_0),\\
\nonumber
x\Delta d_v(x,Q^2_0)&=&\eta_d A_dx^{a_d}xd_v(x,Q^2_0),\\
\nonumber
x\Delta s(x,Q^2_0)&=&\eta_s A_sx^{a_s}xs(x,Q^2_0),\\
x\Delta G(x,Q^2_0)&=&\eta_g A_gx^{a_g}xG(x,Q^2_0),
\label{inputPPD}
\end{eqnarray}
where on the RHS of (\ref{inputPPD}) we have used the MRST99
(central gluon) \cite{MRST99} parametrizations for the NLO($\rm
\overline{MS}$) unpolarized densities. The normalization factors
$A_i$ in (\ref{inputPPD}) are fixed such that $\eta_{i}$ are the
first moments of the polarized densities. The first moments of the
valence quark densities $\eta_u$ and $\eta_d$ are constrained by
the baryon decay constants (F+D) and (3F-D) \cite{LSS05} assuming
$\rm SU(3)_f$ symmetry. Bearing in mind that the light quark sea
densities $\Delta\bar{u}$ and $\Delta\bar{d}$ cannot, in
principle, be determined from the present inclusive data (in the
absence of polararized charged current neutrino experiments) we
have adopted the convention of a flavor symmetric sea
\begin{equation}
\Delta u_{sea}=\Delta\bar{u}=\Delta d_{sea}=\Delta\bar{d}= \Delta
s=\Delta\bar{s}. \label{SU3sea}
\end{equation}
Note that this convention only affects the results for the valence
parton densities, but not the results for the strange sea quark
and gluon densities.

In polarized DIS the $Q^2$ range and the accuracy of the data are
much smaller than that in the unpolarized case. That is why, in
all calculations we have used a fixed value of the QCD parameter
$~\Lambda_{\rm \overline{MS}}(n_f=4)=300~{\rm MeV}$, which
corresponds to $~\alpha_s(M^2_{z}) = 0.1175$, as obtained by the
MRST NLO QCD analysis \cite{MRST98} of the world unpolarized data.
This is in excellent agreement with the current world average
$~\alpha_s(M^2_{z}) = 0.1176 \pm 0.002~$ \cite{alsaver}.

\section{Results of analysis}

In this section we will discuss how inclusion of the CLAS proton
and deuteron $g_1/F_1$ data \cite{CLAS06} and the {\it new}
COMPASS data on $A_1^d$ \cite{COMPASS06} influence our previous
results \cite{LSS05} on polarized PD and higher twist obtained
from the NLO QCD fit to the world data \cite{EMC,world,COMPASS05},
before the CLAS and the latest COMPASS data were available.

\subsection{Impact of CLAS data}

The CLAS $\rm EG_1/p,d$ data (633 experimental points) we have
used in our analysis are high-precision data in the following
kinematic region: $\{x\sim 0.1-0.6,~Q^2\sim 1-5~\rm GeV^2,~W>2~\rm
GeV\}$. As the CLAS data are mainly low $Q^2$ data where the role
of HT becomes important, they should help to fix better the higher
twist effects. Indeed, due to the CLAS data, the determination of
HT corrections to the proton and neutron spin structure functions,
$h^p(x)$ and $h^n(x)$, is significantly improved in the CLAS $x$
region, compared to the values of HT obtained from our LSS'05
analysis \cite{LSS05} in which a NLO(${\rm \overline {MS}}$) QCD
approximation for $g_1(x,Q^2)_{\rm LT}$ was used (see Table I).
This effect is illustrated in Fig. 1. One can conclude now that
the HT corrections for the proton target are definitely different
from zero and negative in the $x$ region: 0.1-0.4. Also, including
the CLAS data in the analysis, the HT corrections for the neutron
target are better determined in the $x$ region: 0.2-0.4. Note that
$h^n(x)$ at $x \sim 0.5$ was already fixed very precisely from the
JLab Hall A data on the ratio $g^{(n)}_1/F^{(n)}_1$.
\begin{figure}
\includegraphics[scale=0.75]{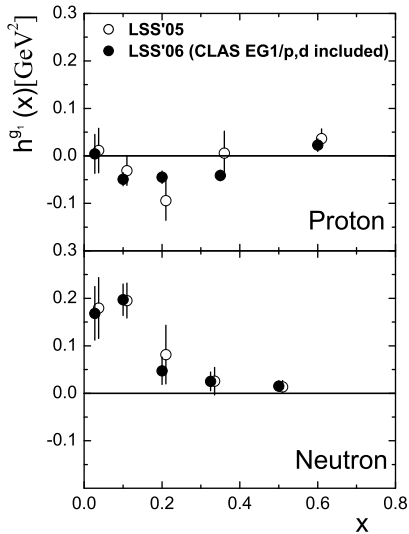}
 \caption{Effect of CLAS data on the higher twist values.
 \label{LSS06_HT}}
\end{figure}

The values obtained for the parameters of the input polarized PD
are presented in Table I and compared with those of LSS'05. Note
that the extracted polarized PD correspond to the Set 2 of
NLO($\rm \overline{MS}$) LSS'05 PPD. As expected, the central
values of the polarized PD are practically {\it not} affected by
the CLAS data (see Table I). This is a consequence of the fact
that at low $Q^2$ the deviation from logarithmic in $Q^2$ pQCD
behaviour of $g_1$ is accounted for by the higher twist term
(\ref{HTQCD}) in $g_1$. Indeed, if one calculates the
$~\chi^2$-probability for the combined world+CLAS data set using
the LSS'05 polarized PD and corresponding HT values, the result
for $\chi^2$ is 938.9 for 823 experimental points, which
significantly decreases to 718.0 after the fit. As seen from Table
I, the best fit to the combined data is achieved mainly through
the changes in the HT values. This supports the theoretical
framework in which the leading twist QCD contribution is
supplemented by higher twist terms of ${\cal O}(\Lambda^2_{\rm
QCD}/Q^2)$. One can see also from Table I, that the accuracy of
the determination of polarized PD is essentially improved. This
improvement (illustrated in Fig. 2) is a consequence of the much
better determination of higher twist contributions to the spin
structure function $g_1$, as discussed above.
\begin{table}
\caption{\label{tab:table1} The parameters of the NLO($\rm
\overline{MS}$) input PPD  at $Q^2=1~GeV^2$ and HT as obtained
from the best fits to the world data \cite{EMC,world,COMPASS05}
(LSS'05) and combined world+CLAS\cite{CLAS06} data set (LSS'06).
The errors shown are total (statistical  and systematic). The
parameters marked by (*) are fixed.}
\begin{ruledtabular}
\begin{tabular}{ccccccc}
    Fit &~~LSS'05 (Set 2)~~&~~LSS'06~~   \\ \hline
 $\rm DF$        &     190~-~16   &  823~-~16  \\
 $\chi^2$        &      154.5     &    718.0    \\
 $\chi^2/\rm DF$ &      0.888     &    0.890    \\  \hline
 $\eta_u$        &    0.926$^*$   &    0.926$^*$  \\
 $a_u$           &~~~~~0.252~$\pm$~0.037~~~~~&~~~~~0.252~$\pm$~0.025~~~~~  \\
 $\eta_d$        &-~0.341$^*$         &    $-0.341^*$       \\
 $a_d$           &~~0.166~$\pm$~0.124~&~~0.166~$\pm$~0.092  \\
 $\eta_s$        &-~0.070~$\pm$~0.008~&-~0.070~$\pm$~0.007  \\
 $a_s$           &~~0.656~$\pm$~0.069~&~~0.679~$\pm$~0.046  \\
 $\eta_g$        &~~0.179~$\pm$~0.267~&~~0.296~$\pm$~0.197  \\
 $a_g$           &~~2.218~$\pm$~1.650 &~~2.465~$\pm$~0.878  \\ \hline
 $x_i$           & \multicolumn{2}{c}{$h^p(x_i)~[GeV^2]$}\\ \hline
  0.028          &~~0.018~$\pm$~0.047 &~~0.004~$\pm$~0.040  \\
  0.100          &-~0.031~$\pm$~0.032 &-~0.049~$\pm$~0.013  \\
  0.200          &-~0.100~$\pm$~0.040 &-~0.045~$\pm$~0.012  \\
  0.350          &~~0.004~$\pm$~0.046 &-~0.041~$\pm$~0.011  \\
  0.600          &~~0.036~$\pm$~0.020 &~~0.023~$\pm$~0.013 \\ \hline
 $x_i$           & \multicolumn{2}{c}{$h^n(x_i)~[GeV^2]$}\\ \hline
  0.028          &~~0.182~$\pm$~0.065 &~~0.168~$\pm$~0.056 \\
  0.100          &~~0.196~$\pm$~0.038 &~~0.197~$\pm$~0.033 \\
  0.200          &~~0.081~$\pm$~0.061 &~~0.047~$\pm$~0.029 \\
  0.325          &~~0.025~$\pm$~0.029 &~~0.025~$\pm$~0.020 \\
  0.500          &~~0.014~$\pm$~0.013 &~~0.015~$\pm$~0.011 \\
\end{tabular}
\end{ruledtabular}
\end{table}
Due to the good accuracy of the CLAS data, one can split the
measured $x$ region of the world+CLAS data set into 7 bins instead
of 5, as used up to now, and therefore, can determine more
precisely the $x$-dependence of the HT corrections to $g_1$. The
numerical results of the best fit to the data using 7 $x$-bins are
listed in Table II (first column). In Fig. 3 the HT values
corresponding to 5 and 7 $x$-bins are compared. As seen in Fig. 3,
the more detailed $x$-space behaviour of the HT contribution,
obtained when using 7 $x$-bins, suggests a smoother function
dependence in $x$ and will help us to calculate more precisely
their first moments in the experimental $x$ region and to compare
them with the predictions given by different models. This point
will be discussed below.
\begin{figure*}
\includegraphics[scale=0.65]{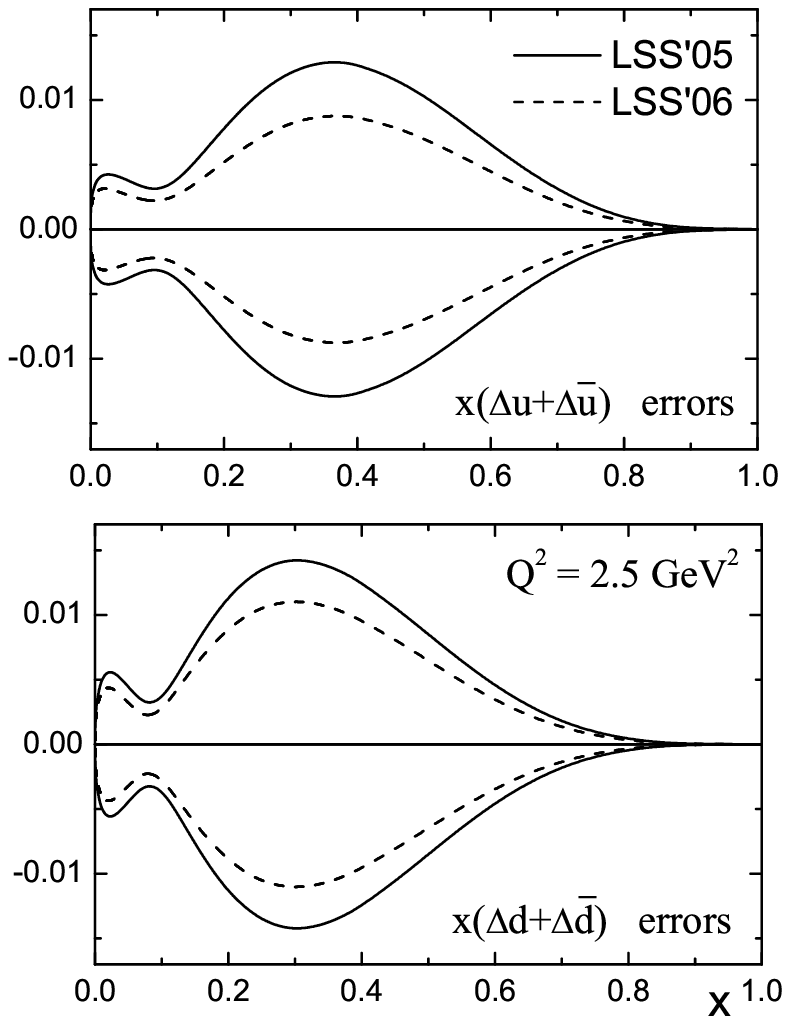}
\includegraphics[scale=0.65]{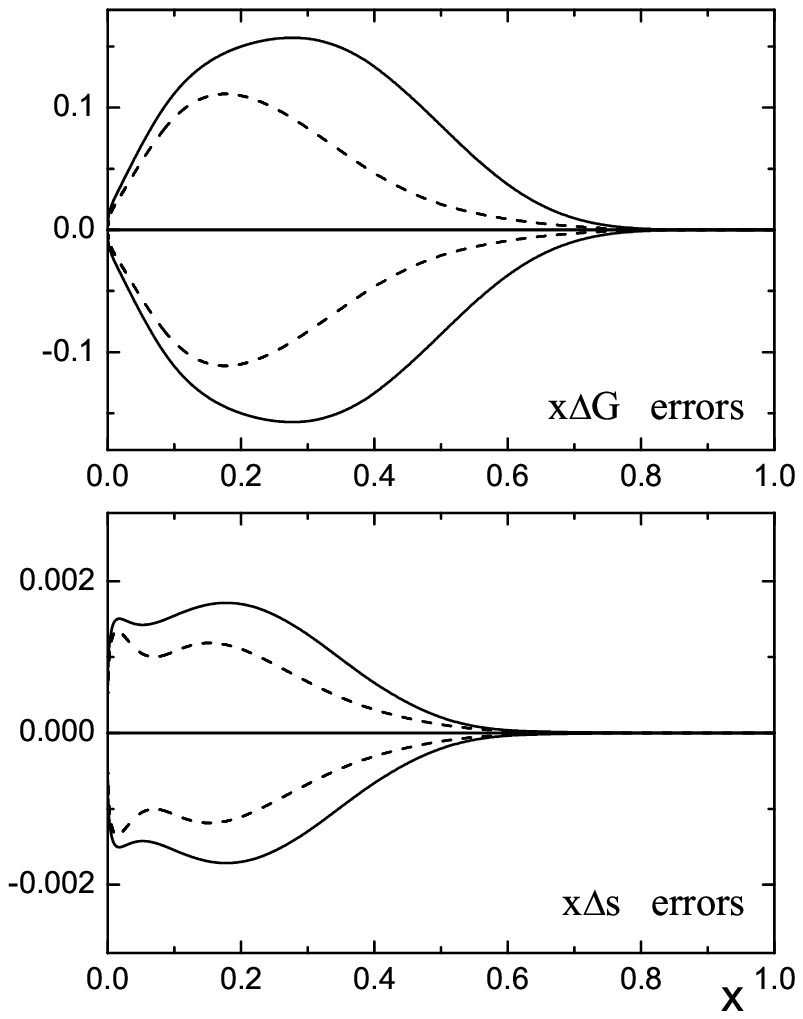}
 \caption{Impact of CLAS data on the uncertainties for
 NLO($\rm \overline{MS}$) polarized quark and gluon densities.
\label{LSS06err}}
\end{figure*}

Comparing the fitted PPD parameters corresponding to 5 and 7
$x$-bins for HT (Table I, 2nd column and Table II, 1st column,
respectively), we observe that the input valence $\Delta u_v$ and
$\Delta d_v$, as well as the strange quark sea $\Delta s$
densities are practically identical. The only exception is the
gluon density, but as seen in Fig. 4, the curve of the gluon
density corresponding to 7 $x$-bins lies within the error band of
$x\Delta G$(5 bins). The curves corresponding to $(\Delta u +
\Delta\bar u)$ and $(\Delta d + \Delta\bar d)$ densities cannot be
distinguished in the experimental region and for that reason they
are not shown in Fig. 4. Note that the curves corresponding to
input $\Delta s(x)$ at $Q^2_0 = 1~GeV^2$ also cannot be
distinguished. They became slightly different at $Q^2 \ne Q^2_0$
(see Fig. 4) because of the mixture between the gluons and sea
quarks due to their $Q^2$ evolution. We consider that the small
correlation between the polarized gluon density and the HT
corrections we have found reflects the fact that the gluons are
not well constrained from the present inclusive DIS data.
\begin{figure}
\includegraphics[scale=0.70]{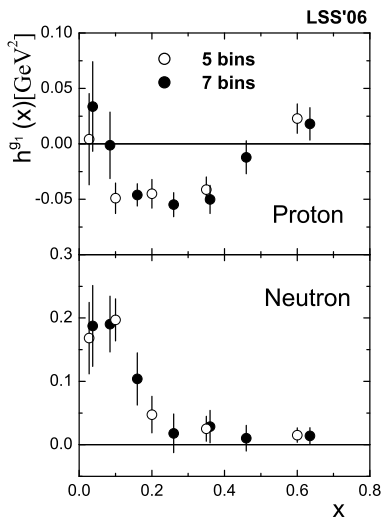}
 \caption{Comparison between the higher twist values corresponding to 5
 and 7 $x$-bins.
 \label{HT5_7bins}}
\end{figure}
\begin{figure}
\includegraphics[scale=0.55]{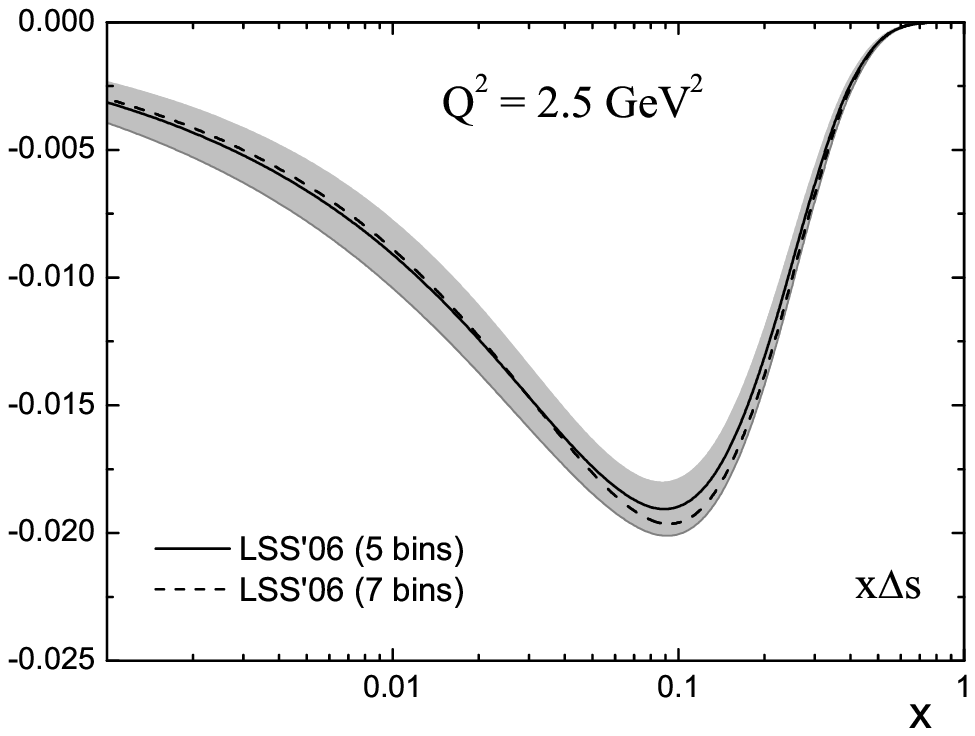}
\includegraphics[scale=0.55]{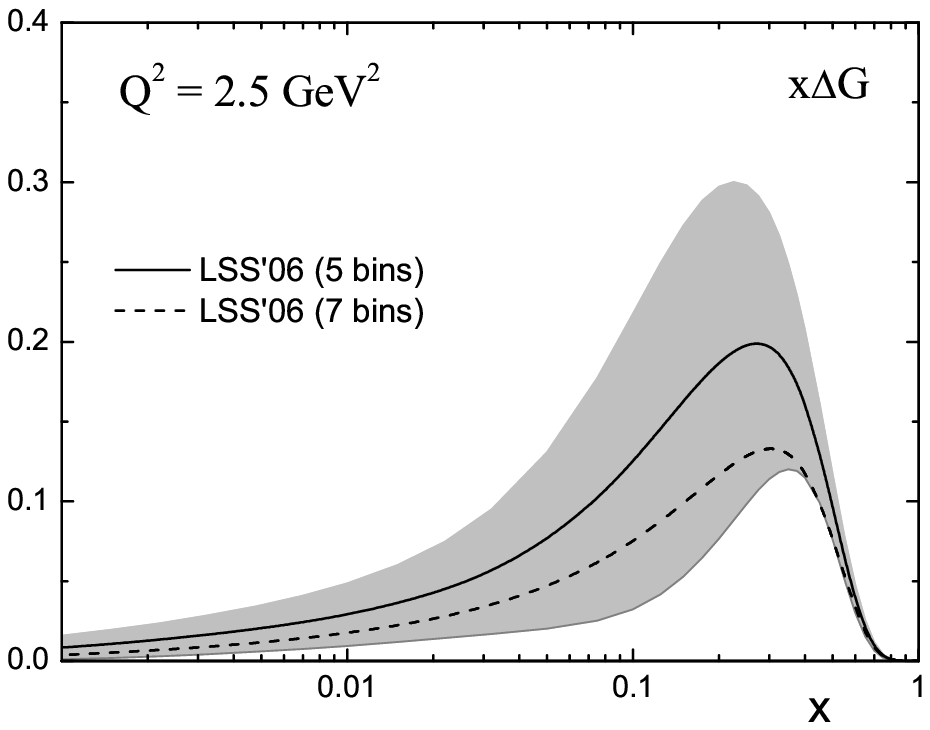}
 \caption{Comparison between polarized NLO($\rm \overline{MS}$) LSS'06 strange
 quark and gluon densities corresponding to fits of the data using 5 and
 7 $x$-bins for higher twist.
\label{PPD5via7bins}}
\end{figure}

\subsection{Impact of new COMPASS data}

When this analysis was finished, the COMPASS Collaboration at CERN
reported new data on the longitudinal asymmetry $A_1^d$
\cite{COMPASS06}. The new data are based on a 2.5 times larger
statistics than those of Ref. \cite{COMPASS05} used in our
analysis. In contrast to the CLAS data, the COMPASS data are at
large $Q^2$ and are the only precise data covering the low $x$
region: $0.004 < x < 0.015$, where the behaviour of the spin
structure function $g_1^d$ should be more sensitive to the sign of
the gluon polarization. Note also, that due to the larger
statistics the latest COMPASS data give more precise and detailed
information about $A^d_1$ and $g^d_1$ in the above experimental
region (see Fig. 5).

In view of this, we have re-analyzed the data superseding the old
set of COMPASS data with the latest one in order to study the
impact of the new COMPASS data on the results reported above. 7
$x$-bins for extracting the HT values were used in the fits. The
numerical results are listed in Table II. As mentioned in Section
II, the logarithmic $Q^2$ dependence of higher twist $h(x, Q^2)$
is neglected in our analysis. So, the numerical values $h(x_i)$
presented in Table II correspond to the mean value of $Q^2$ for
any $x$-bin. Keeping in mind that the higher twist values $h(x_i)$
are mainly determined from the preasymptotic region ($Q^2 \sim
1-5~\rm GeV^2, ~W>2~\rm GeV$), the mean values $<Q^2_i>$ in Table
II correspond to the experimental points for each $x$-bin with
$Q^2 \leq 5~\rm GeV^2$.
\begin{table*}
\caption{\label{tab:table2} Effect of the new COMPASS data on
polarized PD and HT. The parameters of the NLO($\rm
\overline{MS}$) input PPD are given at $Q^2=1~GeV^2$. 7 bins in
$x$ have been used to extract the HT values. The errors shown are
total (statistical and systematic). The parameters marked by (*)
are fixed. }
\begin{ruledtabular}
\begin{tabular}{ccccccc}
   Fit& &LSS'06 ({\it old} COMPASS)& \multicolumn{2}{c} {LSS'06 ({\it new}
   COMPASS)}\\ \hline
 & &$\Delta G > 0$ &~~~$\Delta G > 0$~~~&~~~$\Delta G < 0$~~~ \\ \hline
 $\rm DF$&        &  823~-~20      &     826~-~20   &  826~-~20    \\
 $\chi^2$&        &   716.2        &      721.7     &    722.9      \\
 $\chi^2/\rm DF$& &   0.892        &      0.895     &    0.897       \\  \hline
 $\eta_u$&        &    0.926$^*$   &    0.926$^*$  &~~0.926$^*$ \\
 $a_u$   &        &~~0.264~$\pm$~0.027 &~~~~0.273~$\pm$~0.028~&~~~~0.273~$\pm$~0.028  \\
 $\eta_d$&        &-~0.341$^*$         &    $-0.341^*$      &   $-0.341^*$ \\
 $a_d$   &        &~~0.172~$\pm$~0.118 &~~0.202~$\pm$~0.118~&~~0.160~$\pm$~0.108   \\
 $\eta_s$&        &-~0.070~$\pm$~0.006 &-~0.063~$\pm$~0.005~&-~0.057~$\pm$~0.010   \\
 $a_s$   &        &~~0.674~$\pm$~0.053 &~~0.715~$\pm$~0.052~&~~0.746~$\pm$~0.088   \\
 $\eta_g$&        &~~0.173~$\pm$~0.184 &~~0.129~$\pm$~0.166~&-~0.200~$\pm$~0.414   \\
 $a_g$   &        &~~2.969~$\pm$~1.437 &~~3.265~$\pm$~1.668 &~~0.698~$\pm$~0.806   \\ \hline
 $x_i$ & $<Q^2_i>$ & \multicolumn{3}{c}{$h^p(x_i)~[GeV^2]$}  \\ \hline
  0.028 & 2.0     &~~0.034~$\pm$~0.040 &~~0.010~$\pm$~0.039 &~~0.017~$\pm$~0.041  \\
  0.075 & 2.4     &-~0.001~$\pm$~0.030 &-~0.016~$\pm$~0.030 &-~0.019~$\pm$~0.037   \\
  0.150 & 1.7     &-~0.046~$\pm$~0.010 &-~0.050~$\pm$~0.009 &-~0.056~$\pm$~0.018   \\
  0.250 & 1.8     &-~0.055~$\pm$~0.011 &-~0.059~$\pm$~0.010 &-~0.067~$\pm$~0.013  \\
  0.350 & 2.4     &-~0.050~$\pm$~0.013 &-~0.054~$\pm$~0.012 &-~0.060~$\pm$~0.013  \\
  0.450 & 3.2     &-~0.012~$\pm$~0.015 &-~0.016~$\pm$~0.015 &-~0.020~$\pm$~0.015  \\
  0.625 & 4.1     &~~0.018~$\pm$~0.015 &~~0.016~$\pm$~0.015 &~~0.014~$\pm$~0.015  \\   \hline
 $x_i$  & $<Q^2_i>$        & \multicolumn{3}{c}{$h^n(x_i)~[GeV^2]$} \\ \hline
  0.028 & 1.8     &~~0.187~$\pm$~0.064 &~~0.165~$\pm$~0.064 &~~0.180~$\pm$~0.065  \\
  0.075 & 2.4     &~~0.190~$\pm$~0.044 &~~0.173~$\pm$~0.044 &~~0.174~$\pm$~0.049  \\
  0.150 & 1.4     &~~0.104~$\pm$~0.041 &~~0.107~$\pm$~0.039 &~~0.092~$\pm$~0.040  \\
  0.250 & 1.5     &~~0.018~$\pm$~0.030 &~~0.019~$\pm$~0.030 &~~0.006~$\pm$~0.029  \\
  0.350 & 2.2     &~~0.028~$\pm$~0.026 &~~0.031~$\pm$~0.025 &~~0.019~$\pm$~0.023  \\
  0.450 & 3.0     &~~0.010~$\pm$~0.020 &~~0.013~$\pm$~0.020 &~~0.005~$\pm$~0.019   \\
  0.625 & 3.9     &~~0.014~$\pm$~0.013 &~~0.016~$\pm$~0.013 &~~0.012~$\pm$~0.012   \\
\end{tabular}
\end{ruledtabular}
\end{table*}
\begin{figure}
\includegraphics[scale=0.72]{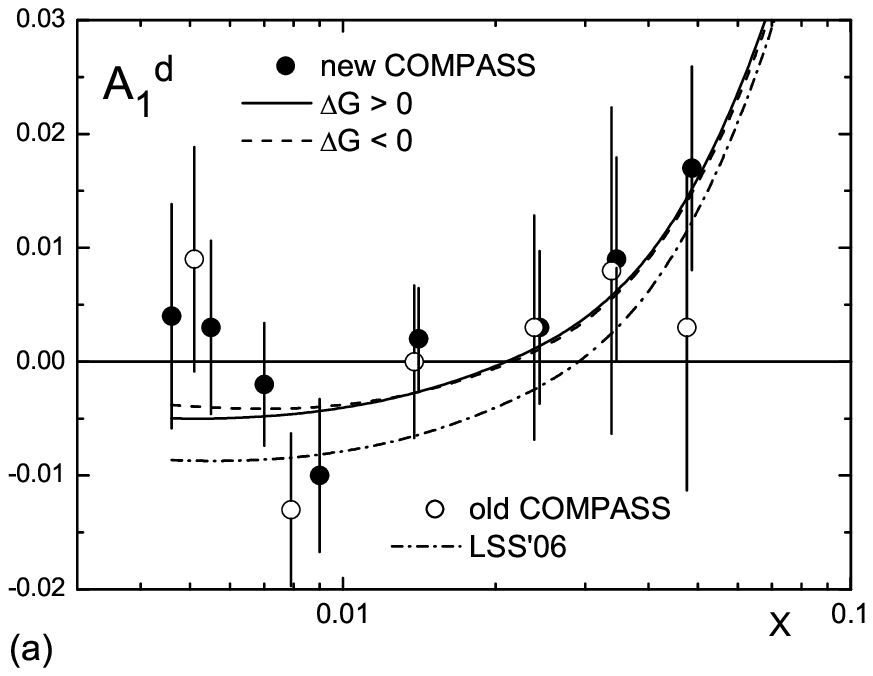}
\includegraphics[scale=0.67]{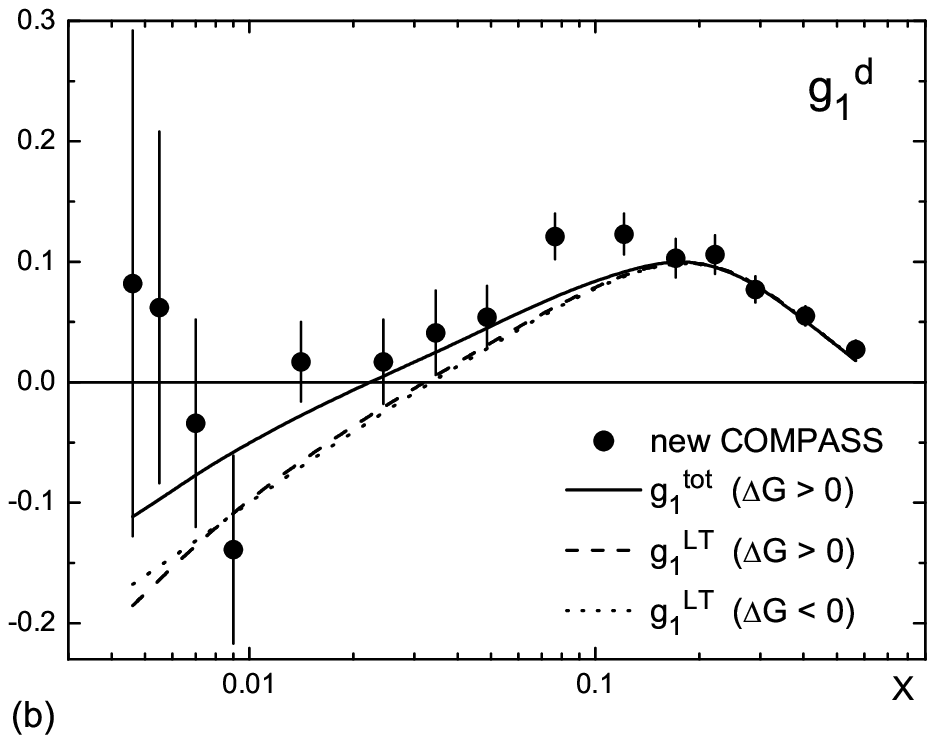}
 \caption{Comparison of our NLO($\rm \overline{MS}$) results for
 $A^d_1$ (a) and $g^d_1$ (b) corresponding to $\Delta G > 0$
 and $\Delta G < 0$ with the new COMPASS data at measured $x$ and
 $Q^2$ values. Error bars represent the total (statistical and
 systematic) errors.
\label{A1g1}}
\end{figure}
\begin{figure}
\includegraphics[scale=0.65]{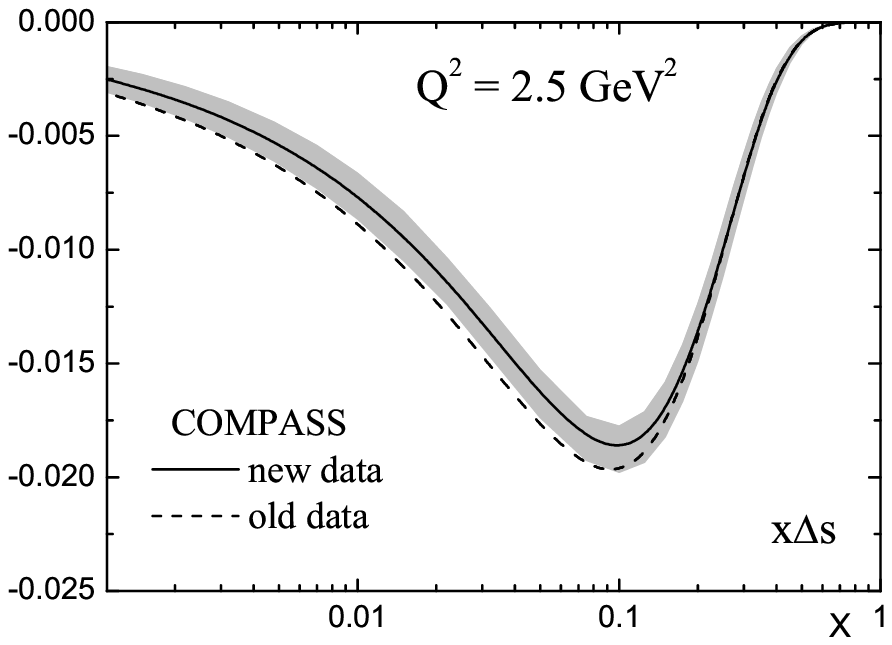}
\includegraphics[scale=0.65]{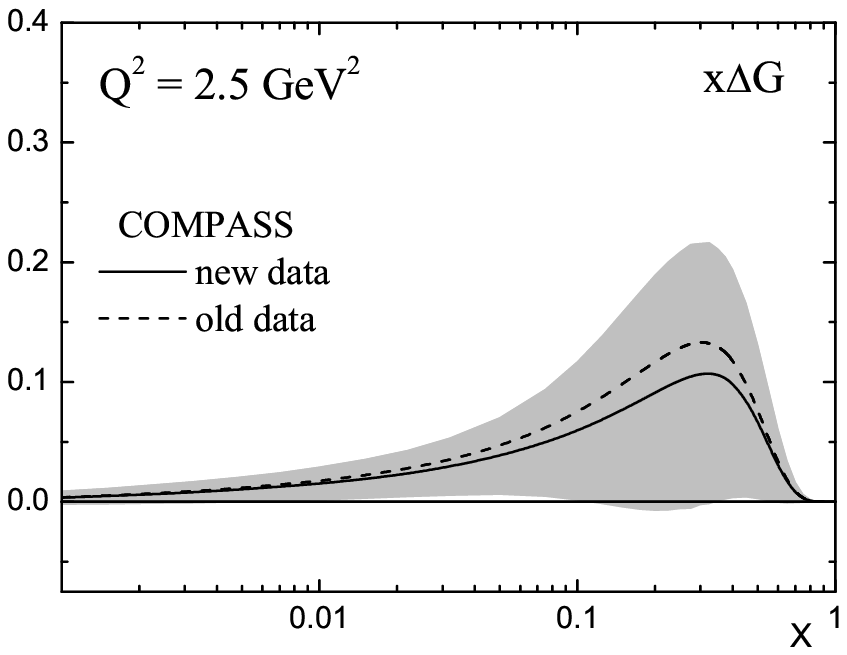}
 \caption{Effect of new COMPASS data on the NLO($\rm \overline{MS}$) LSS'06
 polarized parton densities.
\label{newLSS06}}
\end{figure}

The QCD theoretical curves for $A^d_1$ corresponding to the best
fits to the data with positive and negative gluon polarization are
shown in Fig. 5(a). The old COMPASS data and $A^d_1$ calculated
using the LSS'06 polarized PD and HT corrections (discussed in the
previous Section), are also presented. Note that for $x > 0.1$ the
theoretical curves corresponding to the fits using the new or old
set of the COMPASS data cannot be distinguished and for that
reason this $x$ region is not shown in Fig. 5(a). The best fit to
the new $g_1$ data is illustrated in Fig. 5(b).

The effect of the new data on the polarized parton densities and
the higher twist corrections is illustrated in Fig. 6 and Fig. 7,
respectively. While $(\Delta u + \Delta\bar u)$ and $(\Delta d +
\Delta\bar d)$ parton densities do not change in the experimental
region (for that reason they are not shown in Fig. 6), the
magnitudes of both the polarized gluon and strange quark sea
densities and their first moments slightly decrease (see Fig. 6
and Table II). As a consequence, $\Delta \Sigma(Q^2= 1~\rm GeV^2)$
increases from (0.165 $\pm$ 0.044) to (0.207 $\pm$ 0.040) for
$\Delta G >0$ and (0.243 $\pm$ 0.065) for $\Delta G < 0$ (see
below the discussion about $\Delta G < 0$).

As the COMPASS data are mainly at large $Q^2$, the impact of the
new data on the values of higher twist corrections is negligible,
and as expected, they do not improve the uncertainties of HT. The
new central values practically coincide with the old ones (see
Fig. 7(a)). The only exception are the central values of HT at
small $x$ for both the proton and the neutron targets which are
slightly lower than the old ones. Note that this is the only
region where the COMPASS DIS events are at small $Q^2$: 1-4 $\rm
GeV^2$. As a result, in the small $x$ region two opposite
tendencies occur. In order to make $g^d_1$ consistent with zero
for $x < 0.03$, the HT contribution $h^d=(h^p+h^n)0.925/2$, which
is positive, decreases slightly, while $(g^d_1)_{\rm LT}$, which
is negative, grows slightly due to the smaller negative
contribution of $\Delta s(x, Q^2)$ and the smaller contribution of
$\Delta G(x, Q^2)$, convoluted with its Wilson coefficient
function $\delta C_G(x)$, which is negative in this $x$ range (see
Eq. (\ref{g1partons})).

We have also checked the stability of our results with respect to
a change in $\alpha_s(M^2_{z})$, which in our analysis coincides
with its current world average, as mentioned in Section II. When
$\alpha_s(M^2_{z})$ is varied by one standard deviation $\pm
0.002$, the change of the values of the free parameters is within
their errors. In particular, the change of $\eta_s$ and $\eta_g$,
the first moments of the polarized quark sea and gluon densities,
is smaller than 10\% of their standard deviations.
\begin{figure*}
\includegraphics[scale=0.70]{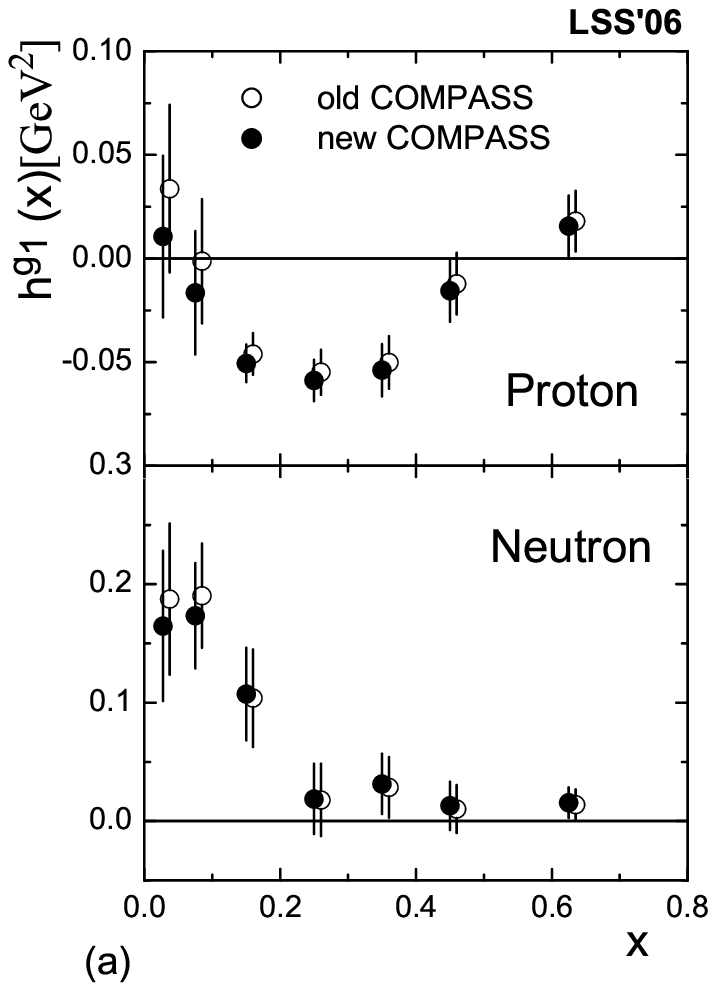}
\includegraphics[scale=0.70]{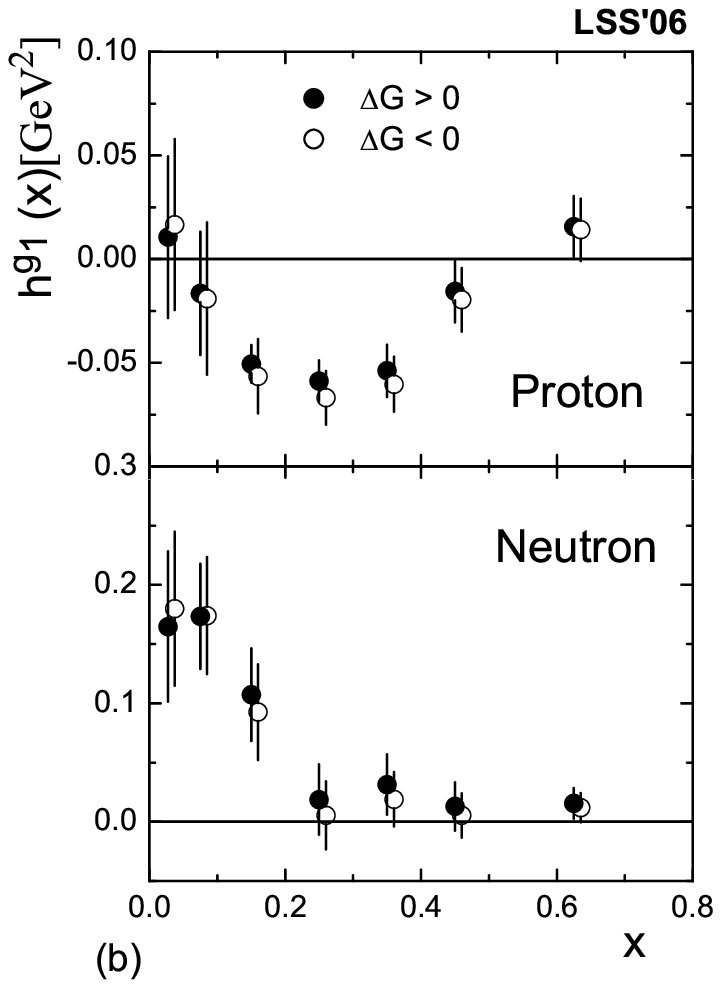}
 \caption{Effect of new COMPASS data on the higher twist values (a).
Comparison between HT values corresponding to the fits with
$\Delta G > 0$ and $\Delta G < 0$ (b).
\label{HT}}
\end{figure*}

\subsection{The sign of the gluon polarization}

We have also studied the possibility of a negative polarized gluon
density. Starting with a negative value for $\eta_g=\Delta
G(Q^2_0)$ (the first moment for the input gluon polarized density
$\Delta G(x, Q^2_0)$), we have found a minimum in $\chi^2$
corresponding to a negative solution for $\eta_g$, and to negative
$\Delta G(Q^2)$ and $x\Delta G(x, Q^2)$. The values of $\chi^2$
corresponding to the fits with $\Delta G > 0$ and $\Delta G < 0$
are practically the same (see Table II) and the data cannot
distinguish between these two solutions for $\Delta G$ (see Fig.
5(a)). Note that in our previous analyses we also found solutions
with negative $\Delta G$, but they were not presented because the
corresponding $\chi^2$ were significantly larger than those
corresponding to the solutions with positive $\Delta G$.

In Fig. 8 the negative polarized gluon density is compared with
the positive one. As seen, the shape of the negative gluon density
differs from that of positive one, but in both cases the magnitude
of $x\Delta G$ is small. Consequently the parton densities
obtained in the fits with $\Delta G > 0$ and $\Delta G < 0$ are
almost identical. For the strange quarks this is illustrated in
Fig. 8. Thus the theoretical curves $(g^d_1)_{\rm LT}$ for the two
types of gluon polarization are practically identical, even in the
region $x < 0.01$ (see Fig. 5(b)).
\begin{figure}
\includegraphics[scale=0.57]{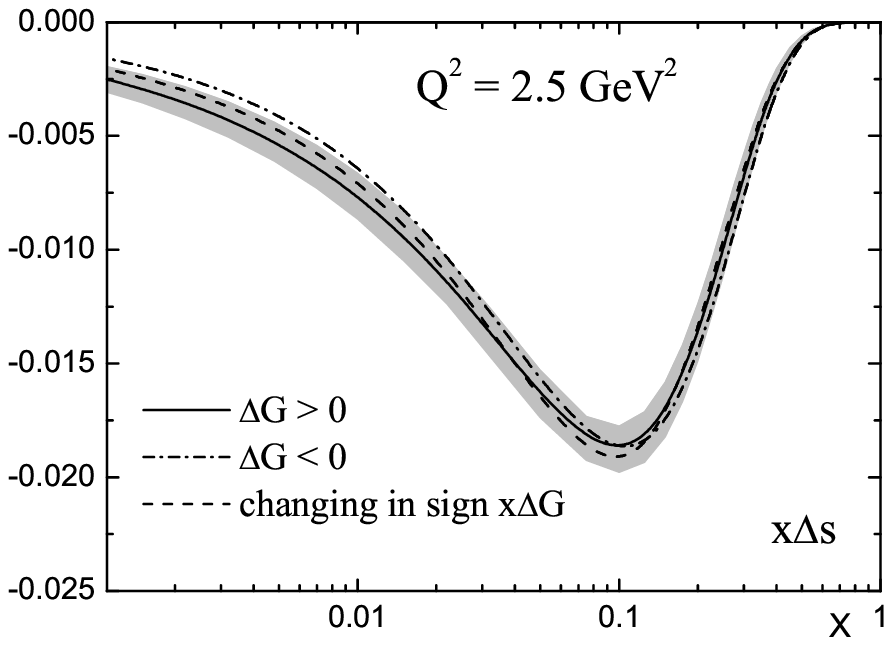}
\includegraphics[scale=0.59]{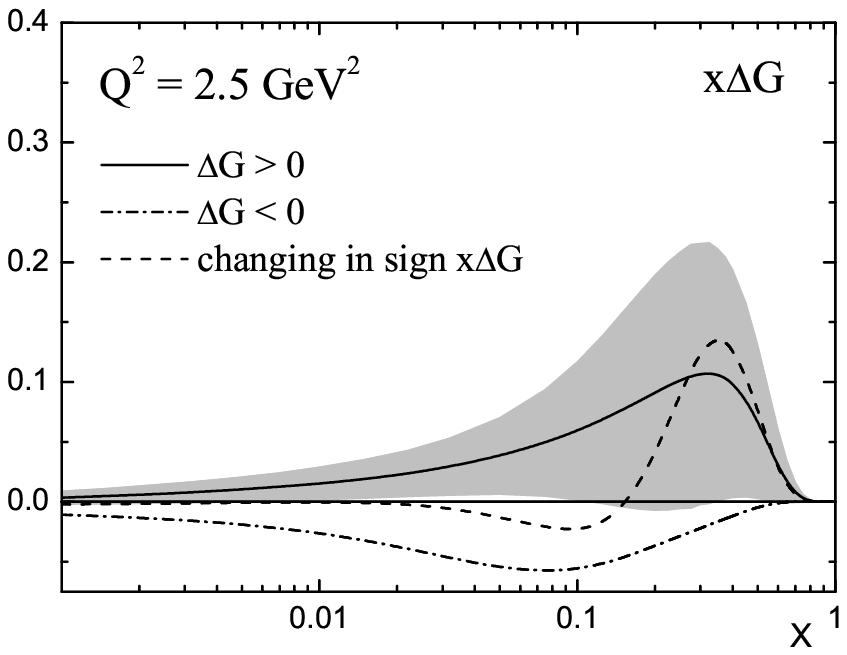}
 \caption{Strange quark sea densities $x\Delta s(x)$ corresponding to the
 fits with $\Delta G > 0$, $\Delta G < 0$ and changing in sign $x\Delta G$.
\label{dsdGpos_neg}}
\end{figure}

Furthermore, as seen in Fig. 7(b), the extracted HT values
corresponding to the fits with $\Delta G > 0$ and $\Delta G < 0$,
are effectively identical. Thus also the total theoretical
expression $(g^d_1)_{\rm tot}$ is essentially the same for $\Delta
G > 0$ and $\Delta G < 0$, even at very small $x < 0.01$.

These results are in contrast to those obtained in the COMPASS
analysis \cite{COMPASS06} where there is a significant difference
between the theoretical curves corresponding to the cases $\Delta
G > 0$ and $\Delta G < 0$ at very small $x$, i.e. in the region
$0.004 <x< 0.02$. The reason for this lies in the question of HT
contributions, which are not taken into account by COMPASS. In the
above $x$ region, $Q^2$ is small ($Q^2 \sim 1-3~GeV^2$) and we
have found that the HT contribution to $(g^d_1)_{\rm tot}$,
$h^d(x)/Q^2$, is positive and large, up to 40\% of the magnitude
of $(g^d_1)_{\rm LT}$ (see Fig. 5(b)). Thus what is fitted by
$(g^d_1)_{\rm LT}(\rm COMPASS)$ is significantly different from
what is fitted by our $(g^d_1)_{\rm LT}(\rm LSS)$ at small $x$,
i.e. $(g^d_1)_{\rm LT}(\rm COMPASS)= (g^d_1)_{\rm LT}(\rm LSS)+
h^d(x)/Q^2$. As a result: {\it i)} The strange quark sea densities
obtained in the two analyses are different, especially in the case
of $\Delta G < 0$ (see Fig. 9 and Fig. 10). {\it ii)} The gluon
densities obtained by COMPASS in both fits ($\Delta G> 0$ and
$\Delta G < 0$) are more peaked than ours.
\begin{figure}
\includegraphics[scale=0.59]{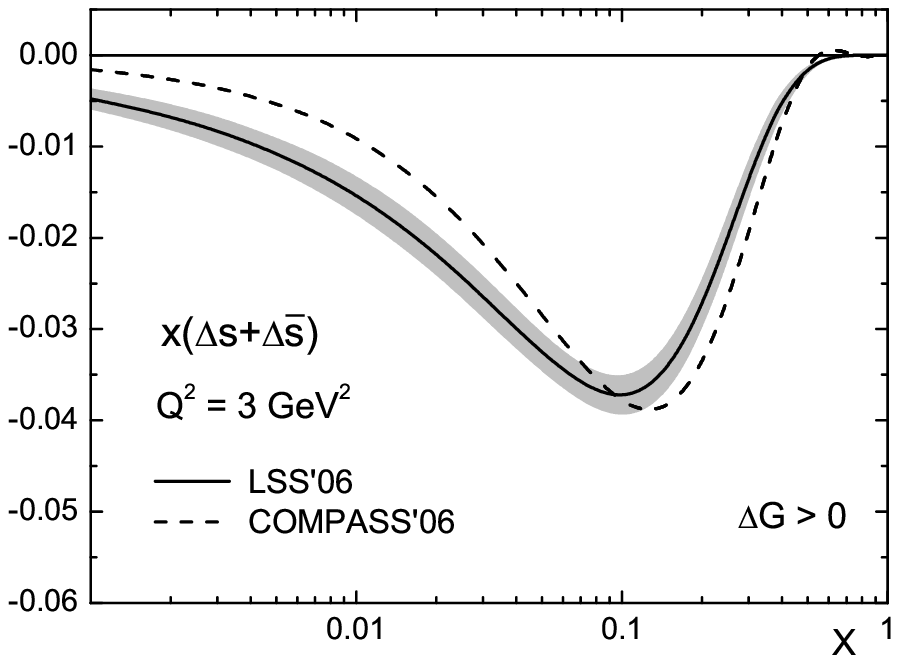}
\includegraphics[scale=0.59]{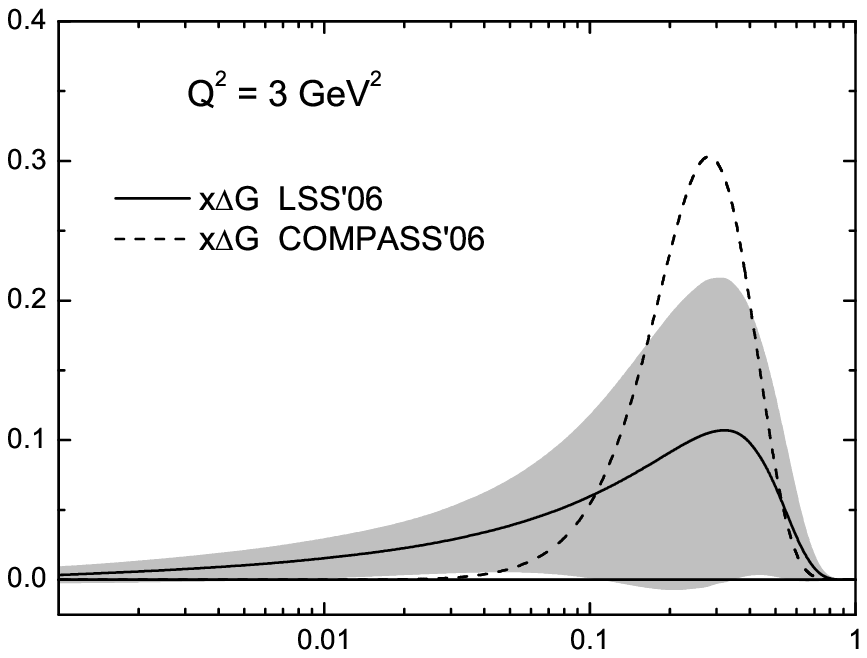}
 \caption{Comparison between our strange quark sea and gluon densities
 corresponding to $\Delta G>0$ and those obtained by COMPASS \cite{COMPASS06}.
\label{LSS06_COMPASS_dgpos}}
\end{figure}
\begin{figure}
\includegraphics[scale=0.59]{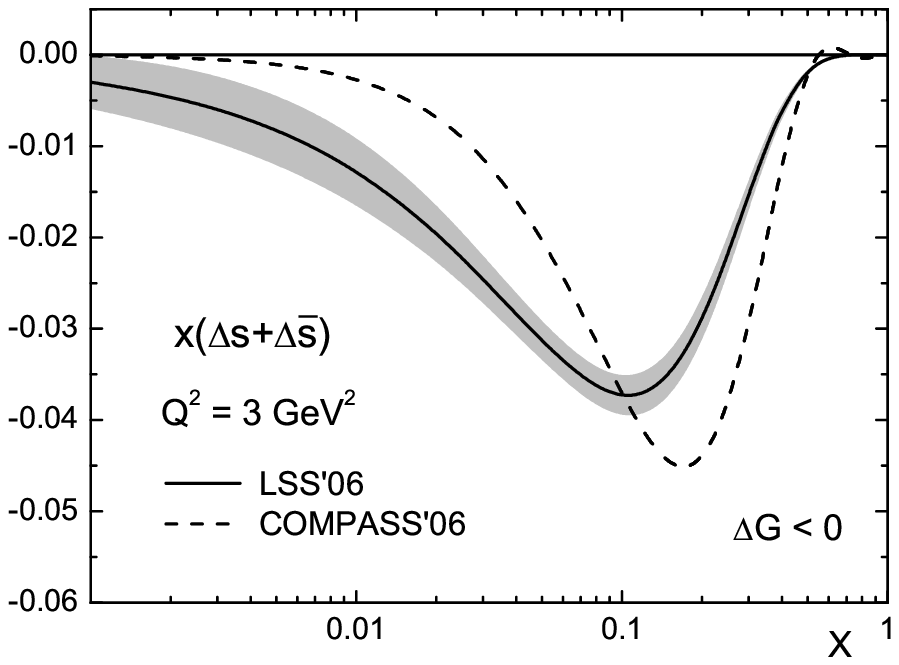}
\includegraphics[scale=0.59]{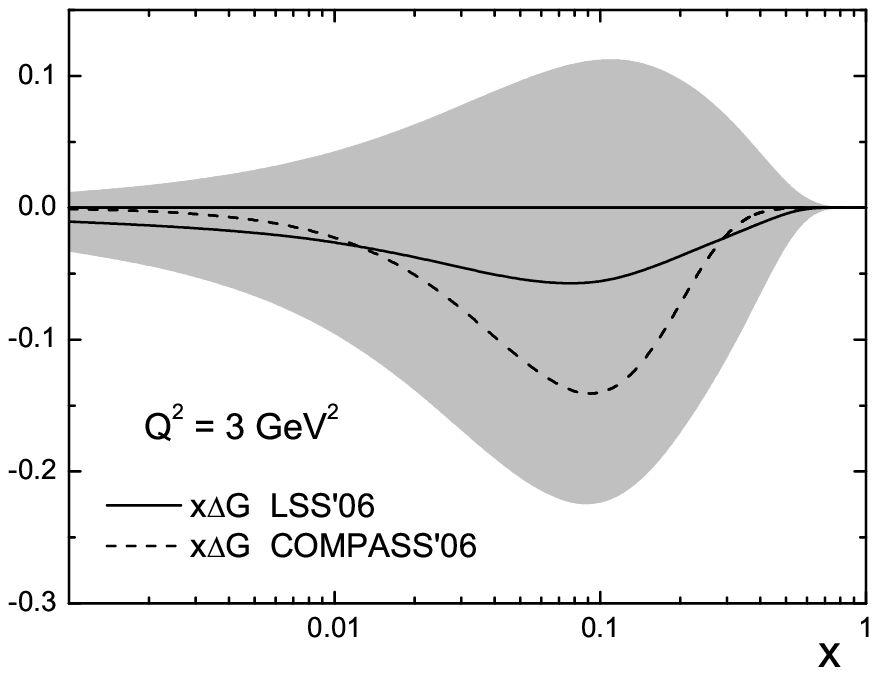}
 \caption{Comparison between our strange quark sea and gluon densities
 corresponding to $\Delta G<0$ and those obtained by COMPASS \cite{COMPASS06}.
\label{LSS06_COMPASS_dgneg}}
\end{figure}

Finally, concerning the possible solution with negative $\Delta G$
we would like to point out the much larger uncertainties in the
determination of the strange quark sea and gluon densities,
$x\Delta s$ and $x\Delta G$, and respectively, their first moments
(see Fig. 11 and Table II). As seen from Fig. 11, the positive
gluon density $x\Delta G(x)$ lies in the error band of the
negative gluons except for $x$ larger than 0.2. $x\Delta s(x)$
corresponding to the positive $\Delta G$ solution lies entirely in
the error band of $x\Delta s(\Delta G<0)$.
\begin{figure}
\includegraphics[scale=0.59]{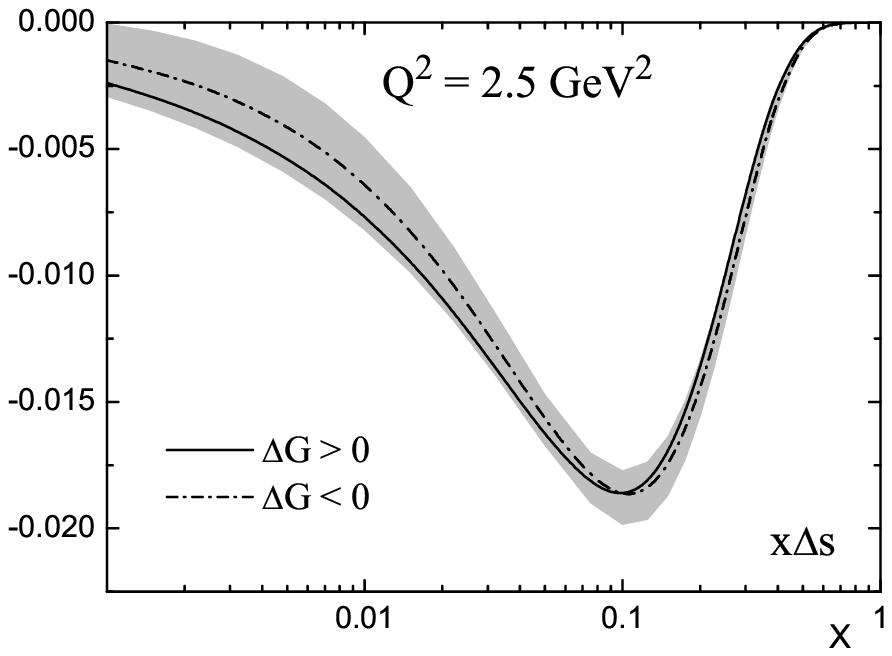}
\includegraphics[scale=0.59]{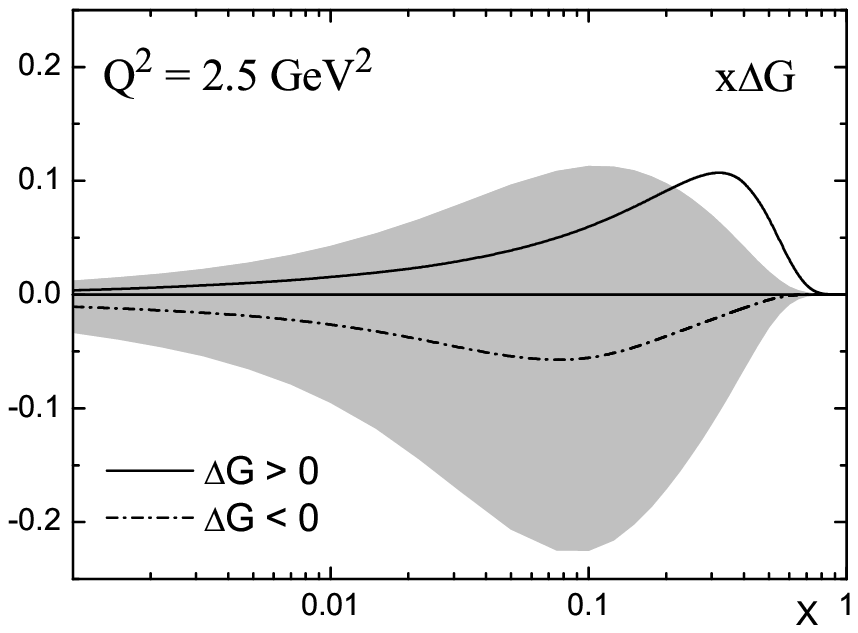}
 \caption{The uncertainties for the strange quark sea and gluon densities
 corresponding to a negative gluon polarization.
\label{errdG_neg}}
\end{figure}

Bearing in mind the high precision of the CLAS and new COMPASS
data over a large range in $Q^2$ we have studied the possibility
to obtain from the fit to the world inclusive DIS data a gluon
density which changes sign as a function of $x$. Such a density
was discussed in \cite{ch_sign_delG} in order to describe the
double longitudinal spin asymmetry $A_{\rm LL}$ of inclusive
$\pi^0$ production in polarized p+p collisions measured by the
{\large P}HENIX \cite{PHENIX} and STAR \cite{STAR} Collaborations
at RHIC. To that end we introduced a factor $~(1+\gamma
x^{\delta})~$  in the input gluon density in (\ref{inputPPD}) with
two new free parameters, $\gamma$ and $\delta$, to be determined
from the fit to the data. In Fig. 8, the determined strange quark
and gluon densities at $Q^2=2.5~\rm GeV^2$ are compared with those
corresponding to the positive and negative $\Delta G$ solutions.
As seen from Fig. 8, the oscillating in sign gluon density lies
between those of positive and negative $\Delta G$. The value of
$\chi^2$ per degree of freedom is 0.895, which coincides with the
values obtained with purely positive or negative $x\Delta G(x)$.

Thus, we are forced to conclude that the accuracy and $Q^2$ range
of the present DIS data is not good enough to discriminate between
these three possibilities. At $Q^2=1~\rm GeV^2$, the shape of the
oscillating in sign polarized gluon density is consistent with
that obtained by the AAC Collaboration from a combined analysis of
DIS (CLAS and new COMPASS not included) and $\pi^0$ asymmetry data
\cite{DISpi0}. Note, however, that compared to the central value
of the first moment $\Delta G_{\rm AAC}=-0.56~\pm ~2.16$ at $Q^2=
1~GeV^2$, presented in \cite{DISpi0}, the central value of our
$\Delta G$ is positive, 0.006, and much smaller in magnitude.
Under evolution in $Q^2$ neither $~\Delta G(Q^2)_{\rm AAC}$, nor
our $~\Delta G(Q^2)$ changes sign, and their magnitudes increase
with increasing of $Q^2$. As a result, the shape of the
corresponding gluon densities for $Q^2 > Q^2_0$ will follow
different tendencies:$x\Delta G(x,Q^2)_{\rm AAC}$ becomes negative
for larger $x$ with increasing of $Q^2$, while our gluon density
for $Q^2 > 6~\rm GeV^2$ is positive for any $x$ in the
experimental region (see Fig. 12).
\begin{figure}
\includegraphics[scale=0.68]{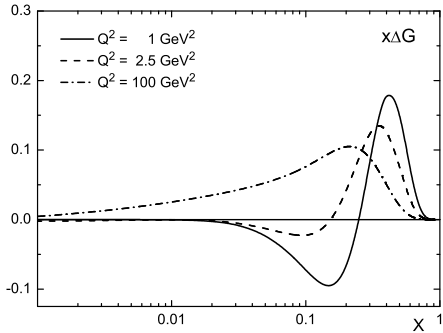}
 \caption{Evolution in $Q^2$ of oscillating-in-sign gluon density.
 \label{chsigndG}}
\end{figure}

In Fig. 13 the ratio $\Delta G(x)/G(x)$ calculated for the
different $\Delta G(x)$ obtained in our analysis and using
$G(x)_{\rm MRST'02}$ taken from \cite{MRST02}, is compared to the
existing direct measurements of $\Delta G/G$ \cite{dG_G_exp}.
(Note that the MRST'02 unpolarized parton densities were used also
in the positivity constraints imposed on the polarized parton
densities obtained in our analysis.) The theoretical curves are
given for $Q^2=3~\rm GeV^2$. The most precise value for $\Delta
G/G$, the COMPASS one, is well consistent with any of the
polarized gluon densities determined in our analysis.
\begin{figure}
\includegraphics[scale=0.72]{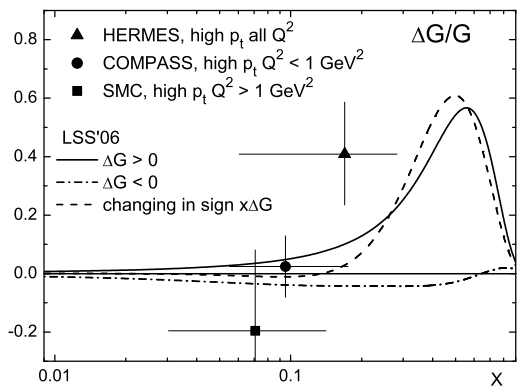}
 \caption{Comparison between the experimental data and
 NLO($\rm \overline{MS}$) curves
 for the gluon polarization $\Delta G(x)/G(x)$ at $Q^2=3~\rm GeV^2$
 corresponding to $\Delta G > 0$, $\Delta G < 0$ and an
 oscillating-in-sign $x\Delta G$. Error bars
 represent the total (statistical and systematic) errors. The
 horizontal bar on each point shows the $x$-range of the measurement.
 \label{delGovG}}
\end{figure}

Finally, let us briefly discuss the values of the first moments of
the higher twist corrections to the proton and neutron structure
function $g_1$. Using the values for $h^N(x)$ corresponding to
$\Delta G>0$ (second column in Table II) we obtain for their first
moments in the experimental region:
\begin{equation}
\bar h^N=\int^{0.75}_{0.0045}h^N(x)dx,~~~~(N=p, n)
\label{momHT}
\end{equation}
$\bar h^p= (- 0.014\pm 0.005)~\rm GeV^2$ for the proton and $\bar
h^n=(0.037 \pm 0.008)~\rm GeV^2$ for the neutron target. As a
result, for the non-singlet $(\bar h^p - \bar h^n)$ and the
singlet $(\bar h^p + \bar h^n)$ we obtain $(- 0.051\pm 0.009)~\rm
GeV^2$ and $(0.023\pm 0.009)~\rm GeV^2$, respectively. The errors
are total (statistical and systematic). The systematic errors are
added qudratically. Note that in our notation
$h=\int^{1}_{0}h(x)dx=4M^2(d_2 + f_2)/9$, where $d_2$ and $f_2$
are the well known quantities, connected with the matrix elements
of twist 3 and twist 4 operators, respectively \cite{d2f2}.

Our values for the first moments for the proton, neutron and
$(\bar h^p - \bar h^n)$ are consistent within the errors with
those extracted directly from the analysis of the first moments of
$g_1^N$ and given in Rfs. \cite{hp}, \cite{hn} and \cite{hp-hn},
respectively. Note that our value for the non-singlet $(\bar h^p -
\bar h^n)$ is in agreement with the QCD sum rule estimates
\cite{Balitsky:1990jb} as well as with the instanton model
predictions \cite{Balla:1997hf,SidWeiss}. The values obtained for
the non-singlet $(\bar h^p - \bar h^n)$ and singlet $(\bar h^p +
\bar h^n)$ quantities are in qualitative agreement with the
relation $|h^p + h^n| << |h^p - h^n|$ derived in the large $\rm
N_c$ limit in QCD \cite{Balla:1997hf}.

\section{Summary}

We have studied the impact of the CLAS and latest COMPASS data on
the polarized parton densities and higher twist contributions. It
was demonstrated that the inclusion of the low $Q^2$ CLAS data in
the NLO QCD analysis of the world DIS data improves essentially
our knowledge of higher twist corrections to the spin structure
function $g_1$. As a consequence, the uncertainties in the
longitudinal polarized parton densities become smaller. The
central values of the densities, however, are not affected and
they practically coincide with those of LSS'05 polarized parton
densities determined from our previous analysis. In contrast to
the CLAS data, the new more precise COMPASS data influence the
strange quark density, but practically do not change the HT
corrections. Given that the COMPASS data is mainly at large $Q^2$,
this behaviour supports the QCD framework, in which the leading
twist pQCD contribution is supplemented by higher twist terms of
${\cal O}(\Lambda^2_{\rm QCD}/Q^2)$.

We have observed that the fit to the world $g_1$ data involving
the CLAS and new COMPASS data yields three possible solutions for
the polarized gluon density, $\Delta G(x) > 0,~\Delta G(x) < 0$
and an changing-in-sign $\Delta G(x)$, which equally well describe
the present DIS data. Also, all of them are in a good agreement
with the directly measured quantity $\Delta G(x)/G(x)$ reported by
COMPASS, although their shapes are very different. We have found
that the magnitude of the gluon polarization is small, $|\Delta G|
< 0.3$ at $Q^2= 1~\rm GeV^2$. We have also found that the higher
twist contribution to $g^d_1$ in the $x$ range $0.004 <x< 0.03$ is
positive and large, up to 40\% of the magnitude of $(g^d_1)_{\rm
LT}$ at $\{x=0.0046,~Q^2=1.1~GeV^2\}$ and therefore, $~g^d_1$ is
not too sensitive to the sign of the gluon polarization in the
above $x$ region, when the higher twist corrections are taken into
account.

\begin{acknowledgments}
This research was supported by a UK Royal Society Joint
International Project Grant, the JINR-Bulgaria Collaborative
Grant, and by the RFBR (No 05-01-00992, 05-02-17748, 06-02-16215).
\end{acknowledgments}

\end{document}